\documentclass{osa-article}

\journal{boe}

\articletype{Research Article}

\usepackage{bm}

\begin{document}

\title{Machine learning enabled multiple illumination quantitative optoacoustic oximetry imaging in humans}

\author{Thomas Kirchner\hspace{0.01mm} \href{https://orcid.org/0000-0002-3819-1987}{\includegraphics[scale=0.07]{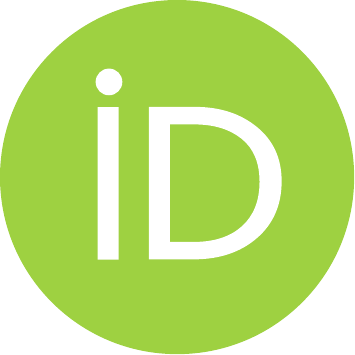}},\authormark{1,2,3} Michael Jaeger,\authormark{2} Martin Frenz\authormark{2,4}}

\address{\authormark{1} Institut für Physik, Martin-Luther-Universität Halle-Wittenberg, Halle (Saale), Germany\\
\authormark{2} Biomedical Photonics, Institute of Applied Physics, University of Bern, Bern, Switzerland\\}

\email{\authormark{3} thomas.kirchner@physik.uni-halle.de}
\email{\authormark{4} frenz@iap.unibe.ch} 


\begin{abstract}
Optoacoustic (OA) imaging is a promising modality for quantifying blood oxygen saturation (sO$_2$) in various biomedical applications -- in diagnosis, monitoring of organ function or even tumor treatment planning. We present an accurate and practically feasible real-time capable method for quantitative imaging of sO$_2$ based on combining multispectral (MS) and multiple illumination (MI)  OA imaging with learned spectral decoloring (LSD). For this purpose we developed a hybrid real-time MI MS OA imaging setup with ultrasound (US) imaging capability; we trained gradient boosting machines on MI spectrally colored absorbed energy spectra generated by generic Monte Carlo simulations, and used the trained models to estimate sO$_2$ on real OA measurements. We validated MI-LSD \emph{in silico} and on \emph{in vivo} image sequences of radial arteries and accompanying veins of five healthy human volunteers. We compared the performance of the method to prior LSD work and conventional linear unmixing. MI-LSD provided highly accurate results \emph{in silico} and consistently plausible results \emph{in vivo}. This preliminary study shows a potentially high applicability of quantitative OA oximetry imaging, using our method.
\end{abstract}

\section{Introduction}

Optoacoustic (OA) imaging (also named photoacoustic imaging) is a promising biomedical imaging modality for visualization of tissue structure and function based on the optical absorption of endogenous or exogenous chromophores.
\cite{xuPhotoacousticImagingBiomedicine2006, xiaPhotoacousticTomographyPrinciples2014}. By combining optical absorption contrast with ultrasonic detection \cite{lauferPhotoacousticImagingPrinciples2018} OA imaging is able to image diagnostically relevant chromophores. Endogenous chromophores like lipids, collagen and melanin can be imaged centimeters-deep in tissue for various clinical purposes. Out of these, preclinically and clinically, OA imaging is most frequently investigated for its ability to image blood. It has been used to image inflammation \cite{knielingMultispectralOptoacousticTomography2017, joPhotoacousticTomographyHuman2018} and other blood volume correlated biomarkers like angiogenesis \cite{siphantoSerialNoninvasivePhotoacoustic2005, horiguchiPilotStudyProstate2017}, but also blood oxygen saturation (sO$_2$) \cite{liPhotoacousticTomographyBlood2018} to assess organ function \cite{kirchnerPhotoacousticsCanImage2019}, wound healing \cite{aizawaPhotoacousticMonitoringBurn2008}, or to follow cancer therapy  \cite{mallidiPhotoacousticImagingCancer2011, suPhotoacousticImagingProstate2011}.

Despite a wide range of approaches, a persistent challenge in OA imaging is the quantification of actual physiological sO$_2$ deep in tissue \cite{coxQuantitativeSpectroscopicPhotoacoustic2012}. 
The main obstacle is the ill-posed inverse problem to the optical transport forward problem. Tissue is a turbid medium with unknown distributions of optical absorption and scattering. Light transport through this medium leads to unknown (illumination geometry dependent and wavelength dependent) optical fluence distributions in the imaged regions. This wavelength dependent optical fluence causes so-called spectral coloring \cite{hochuliEstimatingBloodOxygenation2019, maslovEffectsWavelengthdependentFluence2007} in the measured OA spectra. 
Currently machine learning and data driven approaches are frequently investigated in order to move towards quantitative OA oximetry and imaging in general \cite{grohlDeepLearningBiomedical2021, benchAccurateQuantitativePhotoacoustic2020, tzoumasEigenspectraOptoacousticTomography2016}. Learned spectral decoloring \cite{grohlLearnedSpectralDecoloring2021} is one such approach that aims to estimate sO$_2$ pixel-wise from multispectral (MS) OA signals, with each pixel being evaluated individually -- information from other pixels is not taken into account for the estimation. This tries to breaks down spectral coloring which is effected by a large region, to much fewer dimensions.
In order to add information and constraints to this ill-posed problem, physical additions to the OA sensing apparatus have been investigated. These include diffuse tomography \cite{bauerQuantitativePhotoacousticImaging2011, ulrichSpectralCorrectionHandheld2019} or multiple illumination geometries. \cite{zempQuantitativePhotoacousticTomography2010, shaoEstimatingOpticalAbsorption2011, heldMultipleIrradiationSensing2016, kimCorrectionWavelengthdependentLaser2020} 

Here, we propose that the application of machine learning to local spectra alone is insufficient to perform quantitative OA oximetry -- the ill-posed inverse problem remains (too) ill-posed to arrive at accurate or robust estimates of sO$_2$ without further constraints. We therefore propose to incorporate additional information which may allow us to more accurately estimate sO$_2$. This is done by a combination of multiple illumination (MI) sensing and multispectral sensing with a modified LSD technique.

In earlier work \cite{kirchnerMultipleIlluminationLearned2021} we validated such an approach on a highly controlled copper and nickel sulfate \cite{fonsecaSulfatesChromophoresMultiwavelength2017} phantom model of sO$_2$. In this work we: 

\begin{enumerate}
    \item Developed a hybrid real-time MI MS OA imaging setup with ultrasound (US) imaging capability for human applications.
    \item Designed and performed Monte Carlo simulations of generic human tissue; generating MI spectrally colored absorbed energy spectra.
    \item Designed and trained gradient boosting machines on these simulations and used the trained models to estimate sO$_2$ on \emph{in vivo} human OA measurements. 
\end{enumerate}

We validated MI-LSD \emph{in silico} and on \emph{in vivo} image sequences of radial arteries and accompanying veins of five healthy human volunteers including a hemodynamic challenge on one volunteer. We compared the MI-LSD results to LSD models, conventional linear unmixing of OA data and also pulse oximeter reference measurements. 

\section{Imaging system}

\begin{figure}[b]
\centering\includegraphics[width=9cm]{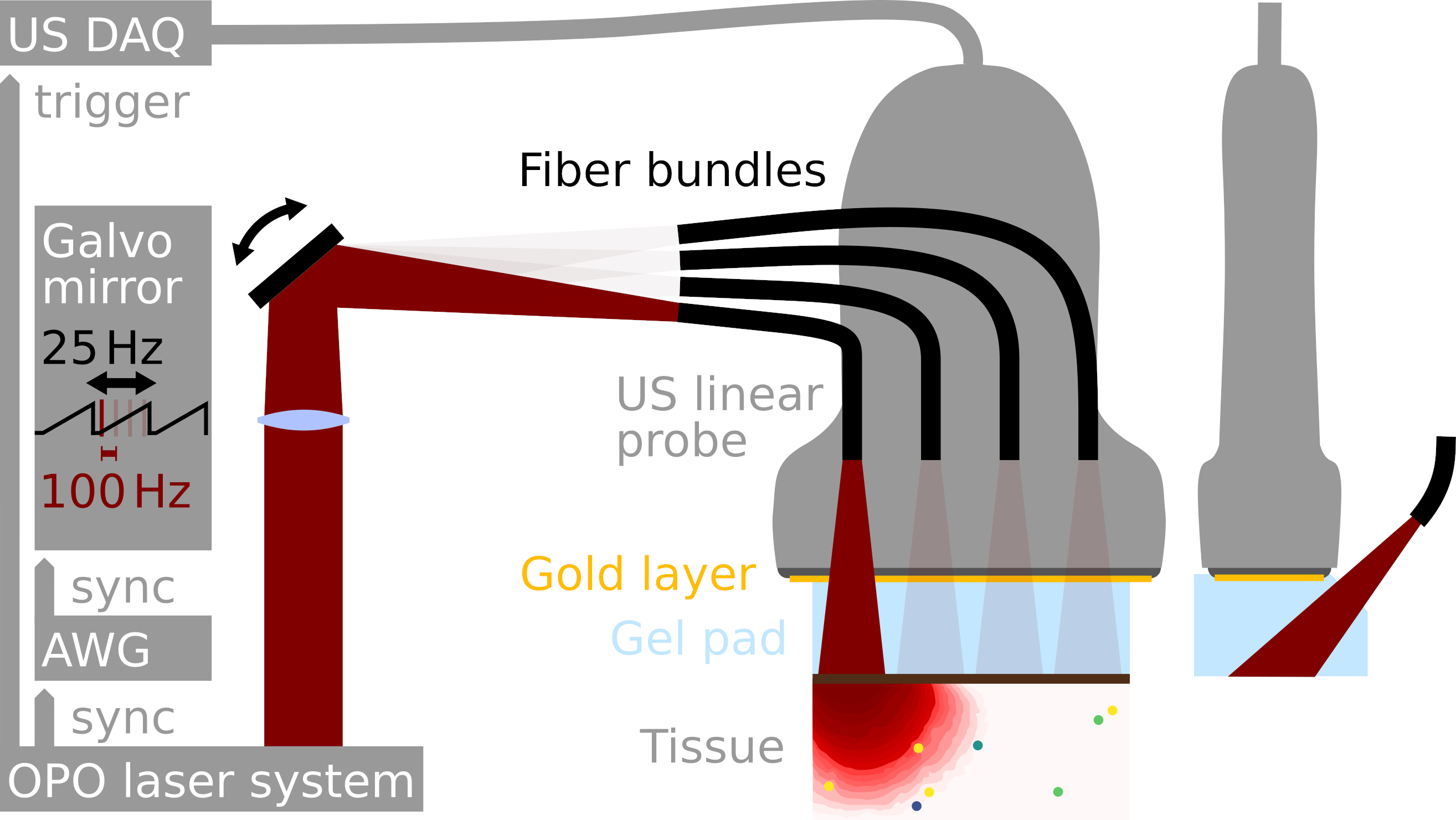}
\caption{Hybrid real-time multiple illumination (MI) optoacoustic (OA) imaging setup with ultrasound (US) imaging. Tissue is illuminated through four fiber bundles by a tunable optical parametric oscillator (OPO) laser system firing laser pulses with 100 Hz. A galvo mirror system is driven at 25 Hz by an arbitrary waveform generator (AWG) to illuminate the fiber bundles in sequence. A linear array US probe detects OA signals which are recorded by a 128-channel US data acquisition (DAQ) system. Conventional US imaging and interleaved OA and US imaging can be performed with the same system. A gold layer is applied to the transducer surface to reduce OA transducer artifacts. A durable copolymer-in-oil gel pad is used to enable in plane illumination.}
\end{figure}

We developed a real-time MI OA imaging system with US imaging capability (shown in figure 1). The system is based on an US data acquisition system (Vantage128, Verasonics, Inc., Kirkland, Washington, USA) with a linear probe (L7-4, Advanced Technology Laboratories Inc., Bothell, Washington, USA) with 128 transducer elements, 5 MHz center frequency, 0.3 mm pitch and 80\% fractional bandwidth. The OA excitation is performed with a fast wavelength-tunable optical parametric oscillator (OPO) laser system (SpitLight, InnoLas Laser GmbH, Krailling, Germany) using 5 ns laser pulses with 100 Hz pulse repetition frequency. Laser pulses are sequentially coupled into four custom fiber bundles (FiberOptic P.+P. AG, Spreitenbach, Switzerland) with NA 0.22 fibers, each bundle 3 mm in diameter. Pulse energy at the fiber exit does not exceed 12 mJ.  The sequential coupling is done with a galvo mirror (GVS011/M, Thorlabs Inc., Newton, New Jersey, USA) and a 25 Hz ramp generated by an arbitrary waveform generator (AWG) (TG5011, Aim-TTi, Cambridgeshire, UK) -- synchronized with the OPO.

The fiber bundle outputs are arranged in a line array with 8 mm spacing. To comply with American National Standards Institute (ANSI) safety limits for skin \cite{ansiAmericanNationalStandard2014}, the beams are widened to 8 mm full-width at half-maximum (FWHM) at the tissue surface, staying below 25 mJ cm$^{-2}$. 
The fiber bundle outputs are attached to the handheld linear probe together with a custom made 10 mm thick gel pad through which illumination and acoustic detection are performed. The gel pad is designed to deliver the illumination pulses to the imaging plane. It is based on a recipe for a copolymer in mineral oil optoacoustic phantom by Hacker et al.~\cite{hackerCopolymerinOilTissueMimickingMaterial2021} -- omitting Titanium Dioxide (TiO$_2$) and Low-density Polyethylene (LDPE) from the phantom recipe to avoid optical scattering in the material. Compared to water based gel pads it can be molded into shape  and fixed to the linear probe much easier while being temporally stable for weeks \cite{hackerCopolymerinOilTissueMimickingMaterial2021}; it also has a higher optical transmittance for some relevant wavelengths as seen in figure 2. The speed of sound in the gel pad was measured to be $\sim$1460 ms$^{-1}$. To reduce the artifact caused by strong optical absorption in the rubber layer of the linear probe, a gold layer (leaf gold $\sim$100 nm thick, \emph{23.75 Karat Blattgold Rosenobel Doppel}) was applied directly to the surface of the linear probe.

\begin{figure}[htbp]
\centering\includegraphics[width=9cm]{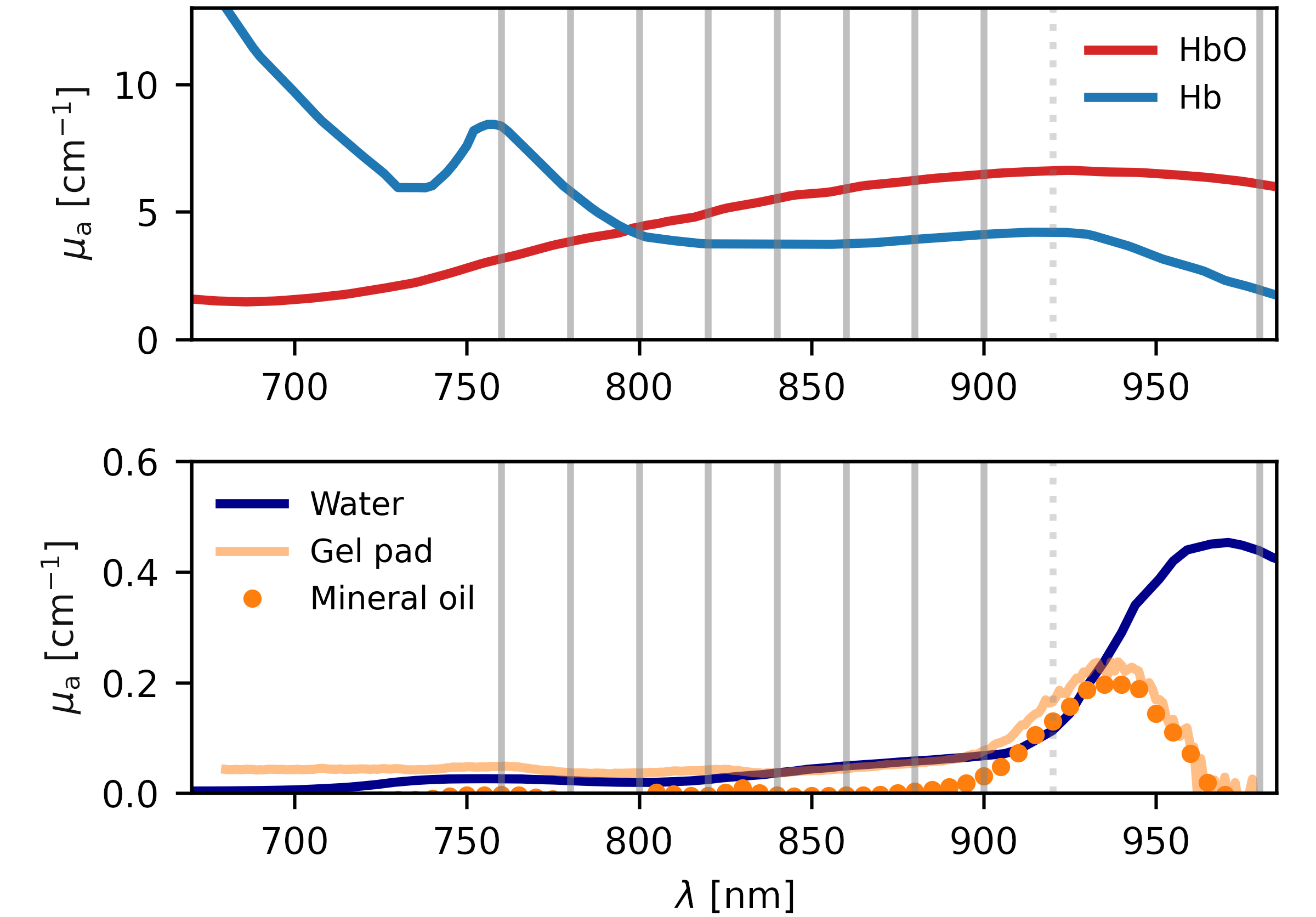}
\caption{Absorption coefficient spectra $\mu_\textrm{a}(\lambda)$ and wavelength selection. Oxy- (HbO) and deoxyhemoglobin (Hb) spectra are shown at whole blood concentrations $c_\textrm{wb}$(Hb or HbO) = 150 gl$^{-1}$ \cite{jacquesOpticalPropertiesBiological2013}. Absorption spectra of a gel pad sample and the base material mineral oil were acquired by transmission spectrometry. Water absorption is shown for reference. Grey vertical lines indicate the wavelengths selected for the multispectral imaging sequence. The dotted line is 920 nm which was part of the recorded wavelength sequence but only intended for an analysis of systematic errors -- see Supplemental Document 1.}
\end{figure}

The MI-MS imaging sequence consisted of four illumination positions, each illuminated by a sequence of ten laser pulses over a range of wavelengths -- (760, 780, 800, 820, 840, 860, 880, 900, 920, 980) nm. While aiming to measure the widest possible range available by the OPO (680 nm to 980 nm), we omitted wavelengths under 760 nm for the following reasons: (1) Variation between pulse energies is highest in the lower tuning range 680--740 nm of the OPO while (2) pulse energy is low, too; (3) Melanin absorption is highest for low wavelengths. Wavelengths between 900 nm and 980 nm were omitted due to the absorption in the gel pad material and the absorption in fat (a similar peak at 930 nm \cite{veenDeterminationVISNIR2004}). This allowed us to reduce the complexity of our simulations by neither simulating absorption of the gel pad material nor fat. 

In addition to the chosen MS sequence -- (760, 780, 800, 820, 840, 860, 880, 900, 980) nm -- we measured at an additional wavelength: 920 nm. This was not used for sO$_2$ estimation in the main body of this work, in the supplement it is used to investigate MI-LSD's robustness to systematic errors in the optical forward model.

One 4 $\times$ 10 pulse sequence is acquired in 0.4 s. For all experiments, we recorded the raw data for 300 such sequences for each subject. Live OA or US beamforming and visualization at 25 fps (showing B-mode US and mean OA signal over all illumination positions) was performed using custom MATLAB scripts. The live visualization was solely used for probe positioning and image quality control (e.g.~avoiding air inclusions in the gel). The following data processing was performed on recorded raw data.

\section{Optoacoustic data processing}

For MI-LSD we train a gradient boosting machine with \emph{in silico} data and estimate sO$_2$ on \emph{in vivo} OA data. An overview of these processes is illustrated in figure 3. The training pipeline and our sO$_2$ estimation method in general is detailed in subsection 3.1 and the \emph{in vivo} data processing pipeline in subsection 3.2. Both pipelines are fully open-source (see Data and Code Availability).

\begin{figure}[htbp]
\centering\includegraphics[width=9cm]{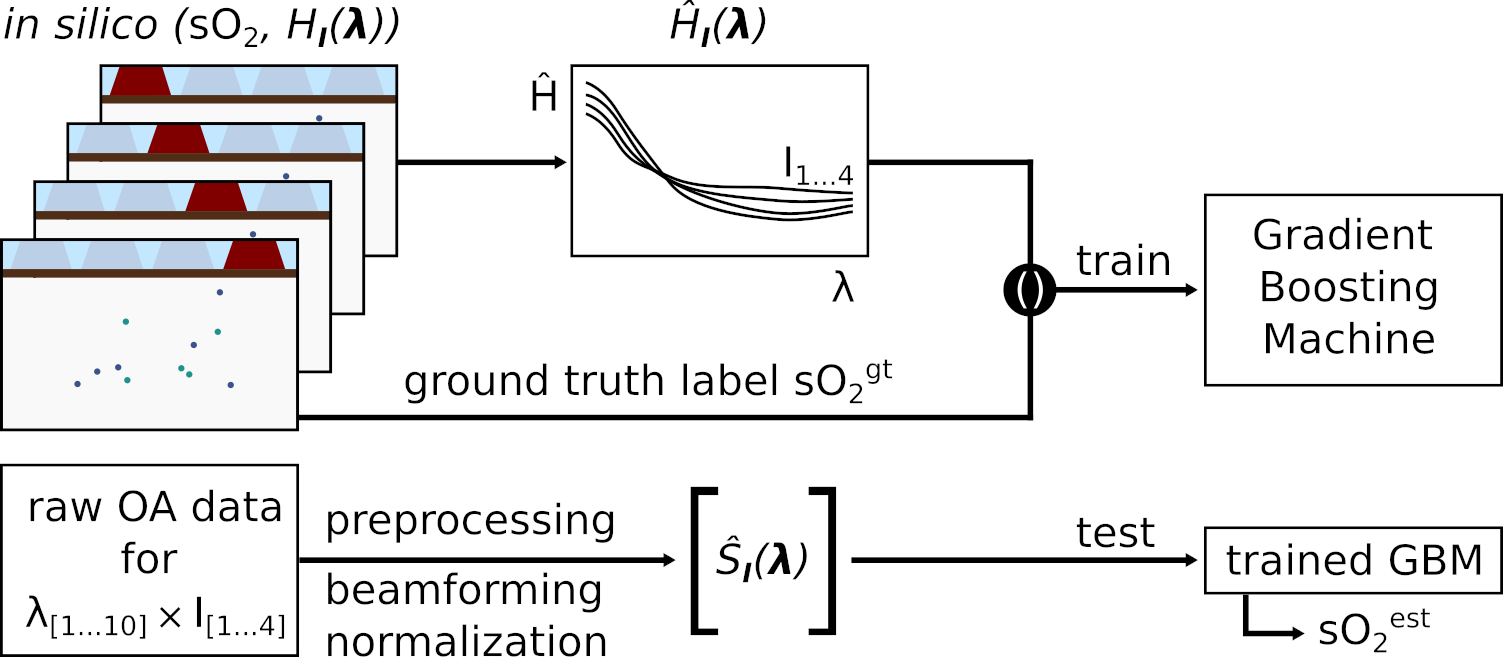}
\caption{Overview of the multiple illumination learned spectral decoloring (MI-LSD) machine learning method using a Gradient Boosting Machine (GBM) algorithm. Monte Carlo simulations of randomized generic tissue volumes generate absorbed energy spectra $H_I(\bm{\lambda})$ for four illuminations $\bm{I}$. The GBM is then trained with tuples of L1 normalized $\hat{H}_I(\bm{\lambda})$ and labels sO$_2^\textrm{gt}$ extracted from these simulations. Raw optoacoustic data is preprocessed and beamformed with a naive delay and sum (DAS) algorithm. L1 normalized signal spectra from these beamformed images $\hat{S}_I(\bm{\lambda})$ are then presented to the trained GBM which estimates sO$_2^\textrm{est}$.}
\end{figure}

\subsection{Blood oxygen saturation estimation method}

Our method is an extension of learned spectral decoloring (LSD) \cite{grohlLearnedSpectralDecoloring2021} and is based on our prior work validating multiple illumination LSD (MI-LSD) on copper and nickel sulfate phantoms \cite{kirchnerMultipleIlluminationLearned2021}. In LSD, we perform a pixel-wise regression on measured spectra $S(\bm{\lambda})$, with $\bm{\lambda}$ being the set of measured wavelengths, in order to estimate sO$_2$ in that pixel. The naive approach to this problem is linear spectral unmixing (LU) \cite{tzoumasSpectralUnmixingTechniques2017, kirchnerOpensourceSoftwarePlatform2019}. In LU $S(\bm{\lambda})$ is fitted to a linear combination of known spectra of known absorbers (i.e. HbO and Hb). In this work, the performed reference LU estimations are based on singular value decomposition using a pseudo inverse absorption matrix. This fast approach was adapted from the simpa toolkit \cite{grohlSIMPAOpenSource2021}.
While LSD aims to estimate sO$_2$ from a single illumination OA signal spectrum $S(\bm{\lambda})$, MI-LSD simply uses multiple such spectra with a set of illuminations $\bm{I}$ as inputs. As in prior work both our LSD and MI-LSD models are machine learning algorithms that are trained on simulated absorbed energy spectra $H(\bm{\lambda})$ labeled with ground truth sO$_2$. These absorbed energy spectra are each normalized with their respective L1 norm resulting into $\hat{H}_I(\bm{\lambda})$; this is done for each illumination $I$ separately. As shown in previous work \cite{kirchnerMultipleIlluminationLearned2021}, this normalization makes them equivalent to normalized measured OA signal spectra $\hat{S}_I(\bm{\lambda})$.

The MI-LSD model is trained on an \emph{in silico} training set consisting of tuples of the absorbed energy spectra  $\hat{H}_I(\bm{\lambda})$ for all four illuminations $\bm{I}$ and the ground truth sO$_2$ label ($\hat{H}_{\bm{I}}(\bm{\lambda})$, sO$_2^\textrm{gt}$). 
The trained model is then presented (1) unseen \emph{in silico} spectra $\hat{H}_{\bm{I}}(\bm{\lambda})$ from a separate test set or (2) OA signal spectra $\hat{S}_{\bm{I}}(\bm{\lambda})$ from an unseen \emph{in vivo} measurement. 
The trained model then estimates the corresponding sO$_2^\textrm{est}$. For the LSD model, illumination is averaged by $\hat{H}_\textrm{LSD} = \sum_{\bm{I}} \hat{H}_I$ and $\hat{S}_\textrm{LSD} = \sum_{\bm{I}} \hat{S}_I$.
As machine learning algorithms we used gradient boosting machine regressors as implemented by the LightGBM framework \cite{keLightGBMHighlyEfficient2017}. In initial experiments on data from previous work \cite{kirchnerMultipleIlluminationLearned2021} they proved faster and less memory intensive than random forests, while providing the same precision. Specifically we used a LGBMRegressor with the \emph{regression\_l1} objective function and 300 estimators (boosting trees) with a maximum of 200 leaves. This hyperparameter tuning was performed exclusively on \emph{in silico} data sets from previous work, and all other hyperparameters were set to default values (also see the open-source code).

To generate \emph{in silico} training and test data sets for the machine learning algorithms we used Monte Carlo simulations of light transport. These simulations were performed with the GPU accelerated open-source mcx toolkit \cite{fangMonteCarloSimulation2009}. The open sourced simpa \cite{grohlSIMPAOpenSource2021} framework, was used for illumination modeling and data management. The aim of these simulations is not to exactly model the specific investigated tissue geometry like in previous work \cite{grohlLearnedSpectralDecoloring2021} but to generate a simplified generic data set that contains the broad range of spectral coloring that is expected for transcutaneous OA imaging.

\begin{figure}[htbp]
\centering\includegraphics[width=9cm]{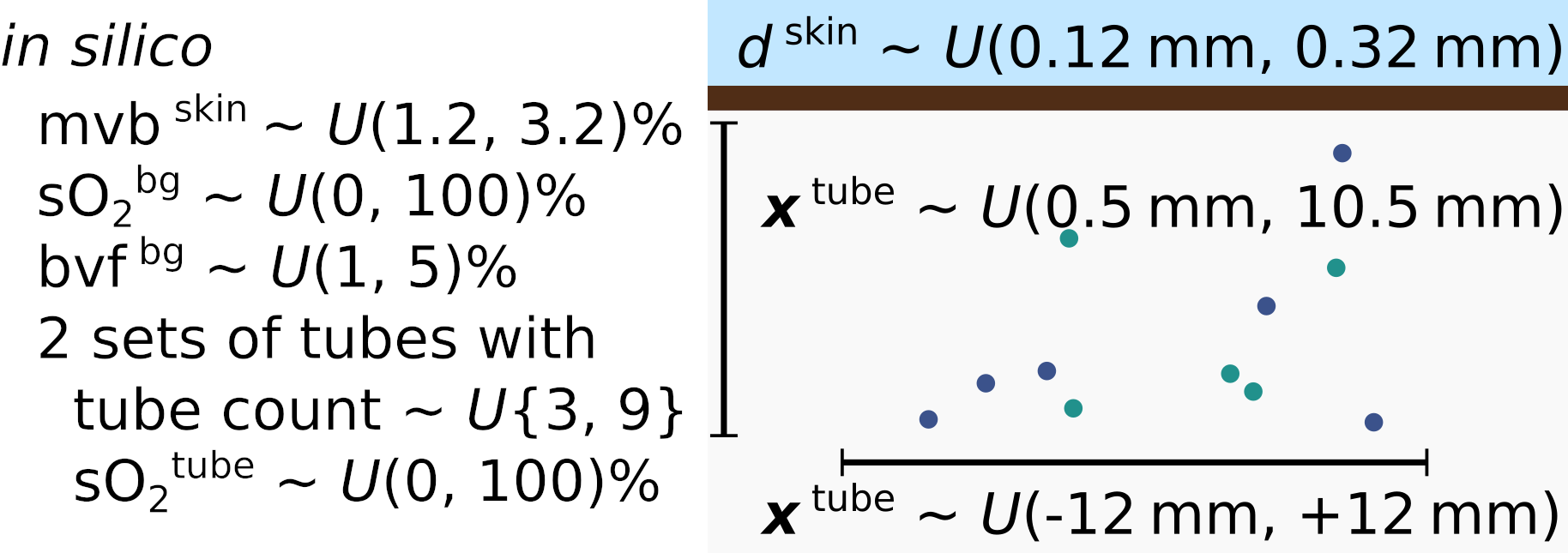}
\caption{Generic tissue geometry used for the Monte Carlo simulations. In order to generate the training set 4000 random tissues are simulated, for the test set 1000 random tissues -- each using the shown parameters. Each volume has two sets of tubes, each set with a random number of tubes, uniformly distributed as specified by $\bm{x}^\textrm{tube}$, and an epidermis layer with thickness $d^\textrm{skin}$. Melanin volume fraction (mvf$^\textrm{skin}$) in the epidermis layer, blood oxygen saturations in the tube set and background tissue (sO$_2^\textrm{bg}$ and sO$_2^\textrm{tube}$) as well as background blood volume fraction (bvf$^\textrm{bg}$) are again drawn from uniform random distributions $U$. The illumination geometry is modeled using the geometry of the setup shown in figure 1.}
\end{figure}

For that we model a broad range of random blood absorbers, background absorption of blood and water, and a randomized epidermis with variable melanin content. The randomization of our tissue layout is shown in figure 4. The \emph{in silico} training data set consists of 4000 volumes and the separate \emph{in silico} test set of 1000 volumes. Each volume was simulated with ten wavelengths and four illumination positions -- mirroring the MI OA imaging sequence with our system. 

Each volume has two sets of tubes with the tube count drawn from a discrete uniform distribution $U$\{3, 9\}. Each tube has a radius of 0.4 mm, is perpendicular to the imaging plane, and is uniform-randomly placed in the volume as specified in figure 4. The sO$_2$ in each tube set and the background sO$_2$ were picked from a continuous uniform random distribution $U$(0\%, 100\%). All tubes within a tube set have the same random sO$_2$. The background blood volume fraction bvf$_\textrm{bg}$ was picked from $U$(0\%, 3\%). The wavelength-dependent background scattering parameters were set using an analytic soft tissue scattering approximation published by Jacques \cite{jacquesOpticalPropertiesBiological2013}. The relevant scattering parameters for background tissue were fixed to average values for soft tissue derived by that meta analysis ($\mu_\textrm{s}'^{500}$ = 19.1, $f^\textrm{ray}$ = 0.153  $b^\textrm{mie}$ = 1.091). For epidermis the scattering was defined using data from Salomatina et al.~\cite{salomatinaOpticalPropertiesNormal2006} ($\mu_\textrm{s}'^{500}$ = 66.7, $f^\textrm{ray}$ = 0.29, $b^\textrm{mie}$ = 0.689). The thickness of the epidermis layer was based on Oltulu et al.~\cite{oltuluMeasurementEpidermisDermis2018} $U$(1.2 mm, 3.2 mm). The melanin volume fraction (mvf) in the epidermis layer was sampled from $U$(1.2\%, 3.2\%) roughly modeling the distribution determined by Alaluf et al.~\cite{alalufEthnicVariationMelanin2002} over all ethnicities for photoexposed and photoprotected skin. For each single wavelength and single illumination position, the Monte Carlo simulation was performed with $2\times10^7$ photon packets. We used 1080 GTX (NVIDIA, Santa Clara) GPUs on a high performance computing cluster. A single-wavelength and single-illumination-position simulation took an average of $\sim$30 s, so that all simulations for the test and training sets used a combined 70 days of GPU time. The absorbed energy density generated by the simulation was averaged over the thickness ($\sim$2 mm) of the experimental imaging plane to increase \emph{in silico} SNR per photon packet. This greatly reduced the number of photons needed for a precise simulation, when compared to previous work. The \emph{in silico} training and test data sets contain only the tube pixels (not the background pixels) and are open data (see the Data and Code Availability Statement).

\subsection{In vivo data processing pipeline}
The recorded raw OA data was preprocessed, including a correction for the mean OPO pulse energy spectrum. This OPO pulse energy spectrum was measured at the fiber bundles outputs before the experiments. For a single wavelength, the variation of pulse energy was $\sim$3\%. For noise and artifact reduction, the data was then bandpass filtered using a Hann window from 0 to 9.6 MHz (corresponding to the frequency range recorded by the DAQ). The beamforming of OA images was performed with a python implementation of delay and sum (DAS) beamforming (using the numba library to increase processing speed). This beamforming implementation was validated against a previous DAS implementation \cite{kirchnerSignedRealTimeDelay2018}. The delay and sum (DAS) algorithm assumed a fixed speed of sound of 1480 ms$^{-1}$. This value was chosen intentionally lower than the typically assumed tissue speed of sound (1540 ms$^{-1}$), to compensate for the lower speed of sound in the 1 cm thick gel pad material (1460 ms$^{-1}$). Furthermore a Hann apodization was used over an apodization angle of $\pm 30$ degrees. The B-mode images were formed using a Hilbert transform filter for envelope detection and then downsampling the result to an isometric resolution of 0.15 mm. See the Data and Code Availability Statement for the open-source image processing scripts used.
Limited by the pulse repetition rate of the laser, one imaging sequence took 0.4 s to measure. On this time scale, tissue motion causes spatial deregistration of the images acquired at different wavelengths, which leads to artifacts in the sO$_2$ estimates. To reduce these unavoidable motion artifacts and also to further increase SNR we averaged our OA B-Mode images over five sequences, as a rolling average.

\section{Experiments and Results}
We tested the performance of MI-LSD for oximetry on human subjects and compared it to LSD and LU sO$_2$ estimation performance. For this we performed (1) an \emph{in silico} validation experiment, (2) an \emph{in vivo} experiment, imaging radial arteries and accompanying veins of five healthy human volunteers and (3) a hemodynamic challenge on one volunteer.

\subsection{in silico}
During the \emph{in silico} validation of the MI-LSD and LSD methods we tested our trained GBM regressors on the \emph{in silico} test set which was kept separately. We compared our results to LU using the same simulated imaging sequence as for LSD. Results are presented for the imaging sequence with the wavelengths (760, 780, 800, 820, 840, 860, 880, 900, 980) nm. 

\begin{figure}[htbp]
\centering\includegraphics[width=13cm]{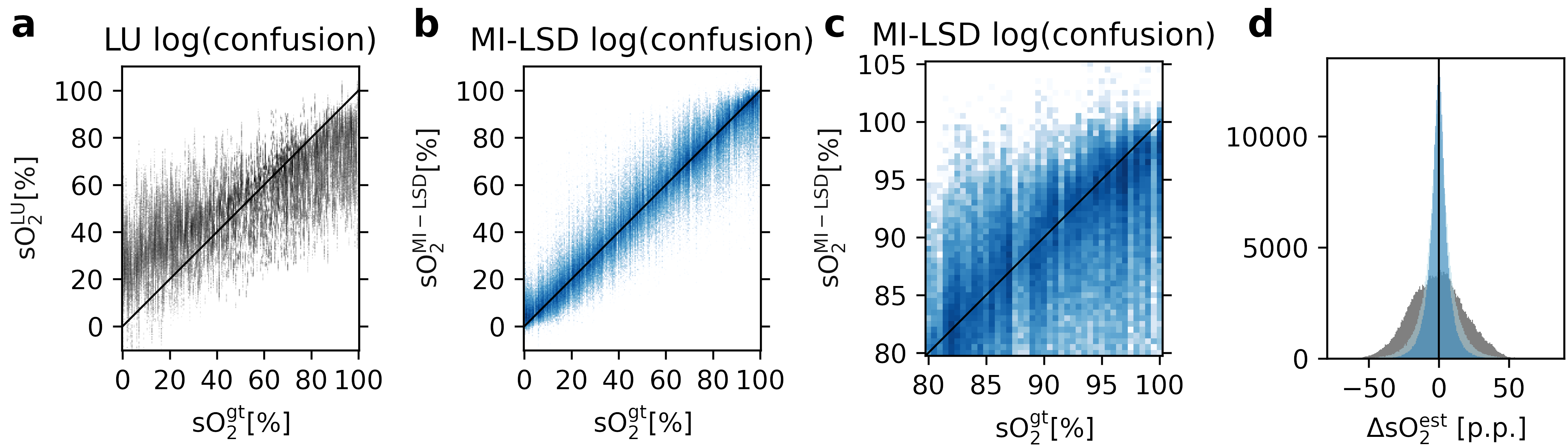}
\caption{Estimates of blood oxygenation saturation (sO$_2$) on the \emph{in silico} test set. Estimated value (sO$_2^\textrm{est}$) with \textbf{a} linear unmixing (LU) and \textbf{b} with multiple illumination learned spectral decoloring (MI-LSD) is plotted against ground truth (sO$_2^\textrm{gt}$) in 2D confusion plots with a logarithmic color scale; in \textbf{c} zoomed in. \textbf{d} shows a histogram over the estimation errors for LU in grey, LSD in light-blue and MI-LSD in blue.}
\end{figure}

The median absolute estimation error $|\Delta \textrm{sO}_2^\textrm{est}|$ \emph{in silico} for MI-LSD was 3.8 percentage points (pp) with an interquartile range IQR of 1.7 pp to 7.4 pp and a 90th percentile P$_{90}$ of 12.6 pp. For LSD the median error was 5.2 pp. with IQR = (2.3 -- 10.7) pp and P$_{90}$ = 18.5 pp. With analytical LU the median error was 13.1 pp. with IQR = (6.1 pp -- 22.3) pp and P$_{90}$ = 31.5 pp.

Regressors were also trained and tested on sequences either including 920 nm illumination or excluding 980 nm (water absorption). The validation results for these modified \emph{in silico} sequences were similar and can be found in the supplemental figures S1 and S2.

The estimation for all 265979 samples in the test set on a high-end consumer CPU (Intel i9-9900KF) was computationally inexpensive with total estimation times of 20 ms for LU, 700 ms for LSD and 800 ms for MI-LSD. Training the GBM regressors on the 1050673 sample training set took 13 s for LSD and 20 s for MI-LSD on the same CPU. 

\subsection{in vivo baseline}
For this experiment we aimed to image easily accessible vasculature with known references for sO$_2$. We therefore imaged the radial artery and surrounding vessels in five subjects. All subjects gave their consent after having been thoroughly informed about the study, and the voluntary nature of participation. Before these measurements on the left forearms of the subjects we took pulse oximeter (OxiMax NPB-40, Nellcor Puritan Bennett Inc., Pleasanton, CA USA) readings (SpO$_2$) from the index finger of the left hand. For each recording of each subject, we imaged a transverse plane of the left forearm eight to twelve centimeters from the wrist. We located the radial artery with the integrated US imaging (see US in figure 6) and positioned the probe such that the radial artery was clearly visible in the center third of the US image -- saving that live US image for reference. We then switched to a live OA view, confirmed that the artery was visible in the OA image and proceeded with a two minute OA raw data recording of the region of interest. This raw data was processed offline using the pipeline described in section 3.2. We then segmented bounding boxes for the radial artery, an accompanying vein and a superficial vessel on the first of the 290 OA images (see boxes for OA in figure 6 ). The vessels were then automatically tracked with fixed size ROIs ($5 \times 3$ pixels) by centering on the maximum intensity OA pixel in each segmented bounding box. All estimates are given as averages over these tracked ROIs. Note that the same regressors trained exclusively on \emph{in silico} data were used for \emph{in vivo} estimation. The LU, MI-LSD and LSD estimates in the pixels of these ROIs were averaged and are shown as time series in figure 6. The time series averages and standard deviations are listed in table 1 together with reference SpO$_2$ readings. Videos showing estimates over the entire image over time for all estimation methods and subjects are part of the open-data supplement. Interleaved US was also recorded for subject 1 and 2. With each of the methods the sO$_2$ estimation can be performed in real time.

\begin{figure}[htbp]
\centering\includegraphics[width=13.3cm]{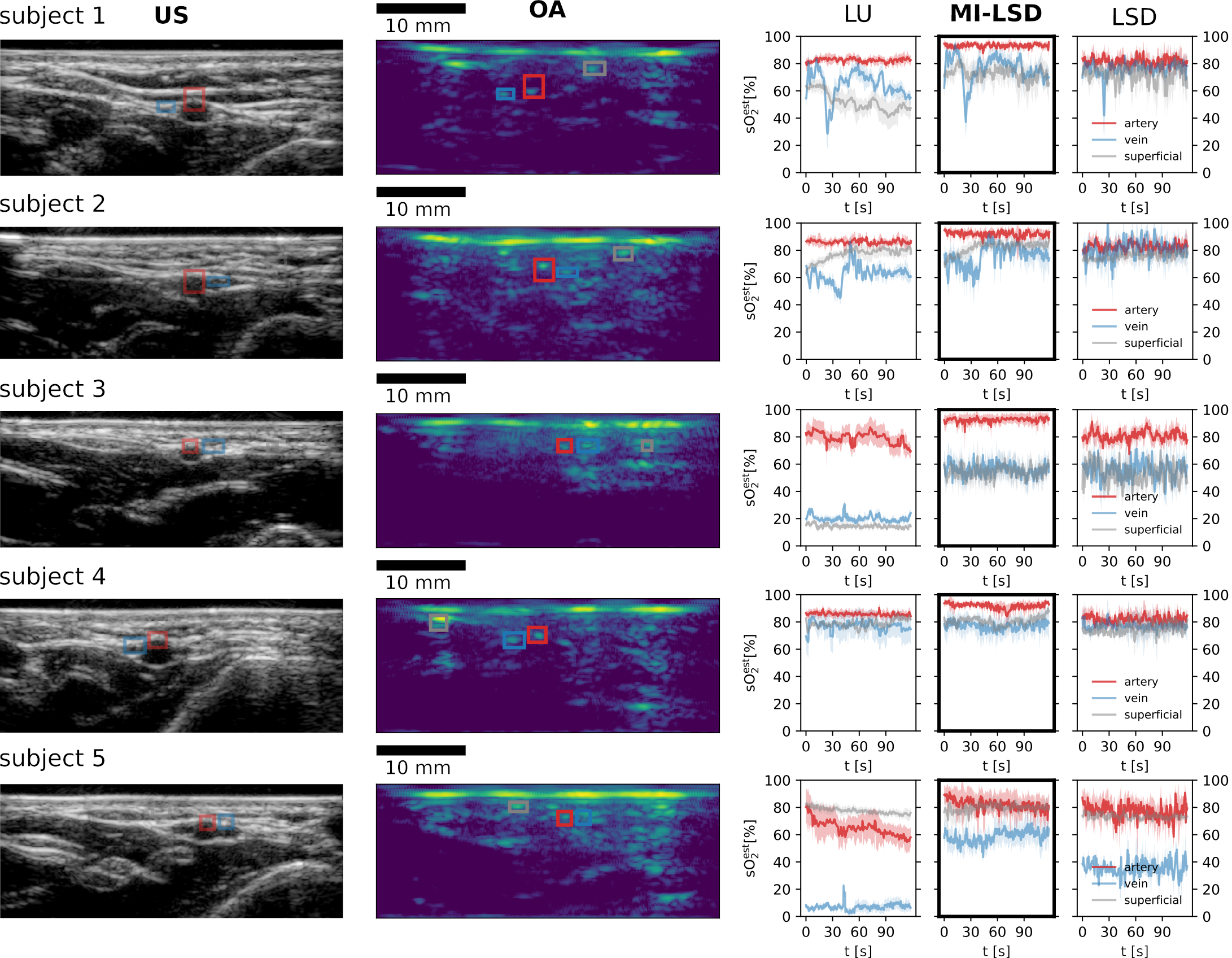}
\caption{Blood oxygen saturation (sO$_2$) estimation on optoacoustic (OA) measurements of the radial artery, radial vein and a superficial vessel in five healthy human volunteers. Ultrasound (US) imaging was used to position the probe and then verify the position of the vessels. Estimation was performed with linear unmixing (LU), multiple illumination learned spectral decoloring (MI-LSD) and mean illumination learned spectral decoloring (LSD), for comparison.}
\end{figure}

\begin{table}[tbh]
\footnotesize
\centering
\begin{tabular}{lllllllllll}
\hline
        & SpO$_2$   & \multicolumn{3}{l}{LU sO$_2^\textrm{est}$} & \multicolumn{3}{l}{\textbf{MI-LSD} sO$_2^\textrm{est}$}  & \multicolumn{3}{l}{LSD sO$_2^\textrm{est}$} \\
        & [\%]    & \multicolumn{3}{l}{mean [\%] $\pm$ sd [pp]} & \multicolumn{3}{l}{mean [\%] $\pm$ sd [pp]} & \multicolumn{3}{l}{mean [\%] $\pm$ sd [pp]} \\
\#      & finger    & artery    & vein      & vessel    & \textbf{artery}    & vein      & vessel    & artery    & vein      & vessel    \\
\hline
1       & 95--97    & $82\pm4$  & $65\pm11$ & $52\pm10$ & $\bm{93 \pm 3}$  & $76 \pm 12$ & $73 \pm 10$ & $81 \pm 7$  & $76 \pm 8$  & $72 \pm 11$ \\
2       & 97        & $86\pm4$  & $63\pm8$  & $77\pm6$  & $\bm{92 \pm 4}$  & $76 \pm 11$ & $82 \pm 6$  & $83 \pm 6$  & $80 \pm 12$ & $76 \pm 6$ \\
3       & 95--98    & $79\pm8$  & $20\pm4$  & $15\pm2$  & $\bm{93 \pm 5}$  & $56 \pm 10$ & $54 \pm 9$  & $81 \pm 7$  & $56 \pm 13$ & $53 \pm 11$ \\
4       & 97        & $86\pm3$  & $79\pm8$  & $79\pm4$  & $\bm{92 \pm 4}$  & $77 \pm 8$  & $80 \pm 5$  & $82 \pm 6$  & $77 \pm 8$  & $75 \pm 7$ \\
5       & 97        & $65\pm11$ & $7\pm4$   & $78\pm4$  & $\bm{82 \pm 10}$ & $59 \pm 10$ & $80 \pm 7$  & $77 \pm 11$ & $36 \pm 9$  & $73 \pm 3$ \\
\hline
\end{tabular}
\caption{Baseline blood oxygen saturation (sO$_2$) estimation on optoacoustic (OA) measurements of the radial artery, radial vein and an additional superficial vessel in five healthy human volunteers. sO$_2$ values are averaged over 290 sequences recorded in two minutes and over a $5\times3$ pixel region of interest (ROI), automatically tracking the vessels. Full sO$_2$  time series are plotted in figure 6. Pulse oximetry (SpO$_2$) measurements on the index finger were taken as a reference before the OA measurement.}
\end{table}

\subsection{in vivo hemodynamic challenge}

\begin{figure}[htbp]
\centering\includegraphics[width=11cm]{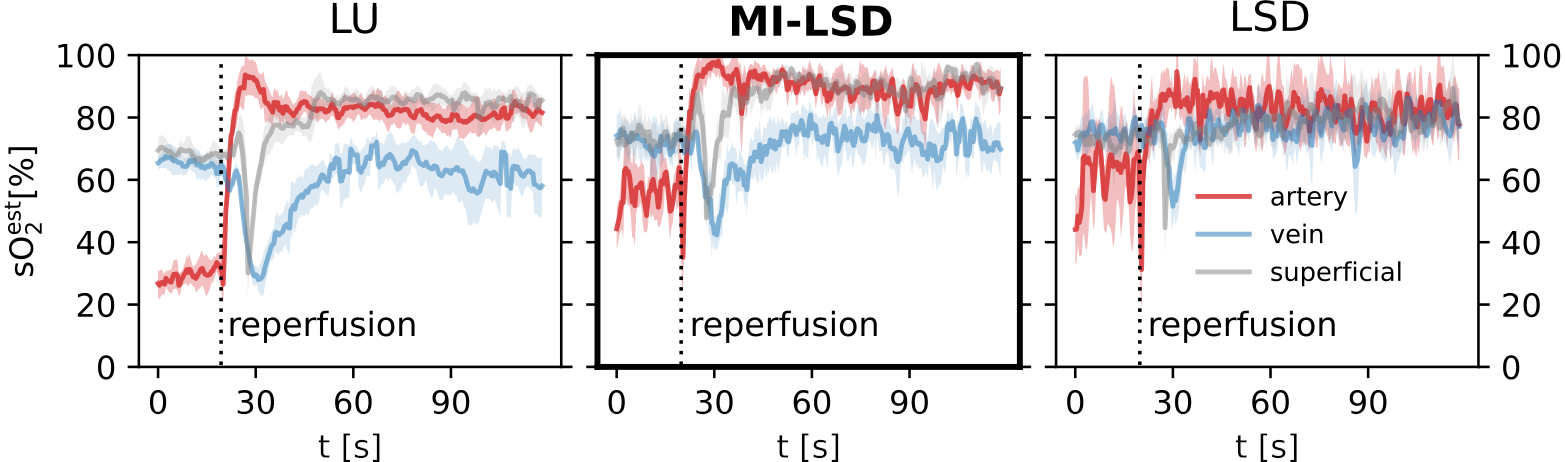}
\caption{Reperfusion experiment. Blood flow was blocked for ten minutes by a blood pressure cuff inflated to 250 mmHg. OA imaging was started 20 seconds before reperfusion by deflation of the cuff. Similar to the baseline experiment, blood oxygen saturation (sO$_2$) estimates are shown for the radial artery, accompanying vein and a superficial vessel. Estimation is performed with linear unmixing (LU), multiple illumination learned spectral decoloring (MI-LSD) and fixed illumination learned spectral decoloring (LSD), for comparison.}
\end{figure}

As an additional experiment we performed a venous and arterial cuff occlusion on the forearm of subject 1. 250 mmHg pressure was applied to occlude arterial and venous blood flow \cite{sudakouTimedomainNIRSSystem2021} for ten minutes. We started OA recording 20 seconds before opening the cuff occlusion and expected a return to baseline for arterial blood after two minutes. All image processing was similar to section 4.2; the resulting time series are shown in figure 7. A pulse oximeter reference measurement could not be taken during this experiment due to the absence of a pulse in the cuffed arm. 

This reperfusion experiment resulted in sO$_2$ estimates with MI-LSD that are consistent with the physiology of such hemodynamic challenges; with MI-LSD showing an estimated arterial sO$_2$ of $56\% \pm 10\%$ before reperfusion (averaged over the first 12 s) and a quick return to baseline with an arterial sO$_2$ of $91\% \pm 4\%$ (averaged over the last 12 s recorded). The arterial sO$_2$ estimates for LSD were $61\% \pm 19\%$ occluded and $82\% \pm 10\%$ reperfused; and for LU, $28\% \pm 4\%$ occluded and $82\% \pm 5\%$ reperfused. 100 seconds after reperfusion, radial artery and vein sO$_2$ returned to subject 1's baseline, and so did the post-experiment pulse oximeter reading.

\section{Discussion and Conclusion}
As expected, the MI-LSD estimations of sO$_2$ on the \emph{in silico} test set were highly accurate, with median absolute estimation errors of 3.8 percentage points, slightly outperforming LSD with $|\Delta \textrm{sO}_2^\textrm{est}|$ = 5.2 pp and clearly outperforming LU with $|\Delta \textrm{sO}_2^\textrm{est}|$ = 13.1 pp. As can be seen in figure 5, MI-LSD (same as LSD and LU) performed worst at the boundaries of 0\% and 100\% sO$_2$. Accuracy of sO$_2$ estimations in the high regime is often medically most relevant as arterial sO$_2$ under 90\% is already considered critical and usually chosen as the clinical threshold for hypoxemia \cite{tylerContinuousMonitoringArterial1985}. The likely reason for this worse performance at the boundary is the non-existence of training samples outside the 0 -- 100\% range which punishes estimating values close to the boundary and leads to an overall bias toward 50\% sO$_2$.

\emph{In silico} accuracy was limited by simulation time, but less by the amount of training data and more by the signal-to-noise of the training data. Simulating more photon packets will likely improve the \emph{in silico} performance but comes at a high computational cost. The MI-LSD model could further be improved by using machine learning algorithms that aim for more explainability \cite{holzingerMachineLearningExplainable2018} of their estimations. This may give us a realistic estimation of accuracy or add confidence estimates to the sO$_2$ estimates of our model \cite{grohlConfidenceEstimationMachine2018}. This would of course also increase computational costs. Using lightGBM is accurate, computationally cheap and greatly reduced memory usage compared to the random forests of our previous work \cite{kirchnerMultipleIlluminationLearned2021}.

Our \emph{in silico} estimation accuracy of MI-LSD is high, even if it is only a slight improvement over LSD. However, this testing on \emph{in silico} data generated with the same random distributions does not imply that MI-LSD or LSD generalizes well. 
The MI-LSD machine learning model was trained by generic and randomized transcutaneous tissue data provided by Monte Carlo simulations. These random data aim to simulate the full range of possible spectral coloring in and by blood and water and include spectral coloring by a wide range of melanin concentration in skin. The aim was to include the full ethnic variation in melanin.
Previous quantitative OA oximetry work \cite{grohlLearnedSpectralDecoloring2021}, when applied to \emph{in vivo} or phantom test data, used \emph{in silico} training data which was not generic but closely modeled after the relevant \emph{in vivo} or phantom test data. In our \emph{in vivo} experiments we estimated sO$_2$ with machine learning models trained with dissimilar and purposely generic \emph{in silico} data. Still, our \emph{in vivo} MI-LSD estimates of sO$_2$ yielded plausible and highly consistent results. The accuracy of these estimates can not be judged well, as these experiments on healthy human volunteers lack hard ground truth data for sO$_2$ which would only be available with invasive means (i.e. arterial blood gas analysis). We employed pulse oximetry to get a non-invasive reference measure for the true arterial sO$_2$. As an additional reference we used literature values for normal arterial sO$_2$ averaging around 97\% \cite{phdGuytonHallTextbook2015}. 

All our arterial blood \emph{in vivo} MI-LSD estimation results for radial arteries are consistent with both literature and pulse oximetry, except for subject 5's. LU and LSD deviate from the  arterial sO$_2$ reference values for all subjects (see table 1). Arterial sO$_2$ baseline estimates for MI-LSD are systematically $\sim$2--5\% lower than expected, which is in line with the deviation observed \emph{in silico}.
For venous blood we only have literature values as references. In central venous blood returning from peripheral tissues sO$_2$ averages around 75\% \cite{phdGuytonHallTextbook2015}, but with wide variation \cite{reinhartValueVenousOximetry2005}. Normal venous sO$_2$ in the arm is usually found to average around 70\% with a large standard deviation of 20\% \cite{belhajComparisonNoninvasivePeripheral2017} or 12\% \cite{keysOxygenSaturationVenous1938}. All our \emph{in vivo} MI-LSD results are consistent with this plausible range for venous blood. LU and LSD have several outliers in their venous sO$_2$ estimates (see table 1 and figure 6).

It is unclear why subject 5 is an outlier in the arterial sO$_2$ estimation using MI-LSD. The falling trendline (see figure 6) could indicate an inadvertent occlusion (maybe partial arterial occlusion) due to pressure applied by the imaging probe. SpO$_2$ was not continuously measured and would have indicated if that were the case. In general, the unequal and uneven pressure applied by the probe to the forearm may also have resulted in a (partial) collapse of the radial veins in some experiments. This may have led to variation over time in sO$_2$ for some of the baselines. 

Generally, motion artifacts also have an additional (blurring) influence on the data. It should also be noted that the acoustic reconstruction with DAS and using a linear transducer array is an imperfect solution to the acoustic inverse problem. This means that we do not in fact estimate from a fully local OA signal spectrum but our OA signal spectra include confounding influences from the surrounding area.
A major limitation of our pilot study is the light skin type in all our subjects. All subjects were light skin males, though with varying degrees of skin photoexposure. The group only included relatively light skin on the ventral side of their forearms (Type 1--3 on the Fitzpatrick scale). Therefore, this pilot study, while promising, yields only limited information about the method's usefulness with the full range of ethnic variation in melanin.
A follow-up study of MI-LSD OA oximetry should include continuous reference measurements with pulse oximetry, possibly a blood gas analysis of venous blood samples and definitely a larger, more diverse cohort.

The result of our hemodynamic challenge experiment showed reduced arterial sO$_2$ estimates for all methods after a ten minute cuff occlusion. Following reperfusion we could observe a characteristic overshoot (hyperemia) of arterial sO$_2$ followed by a return to the baseline of subject 1. Again the MI-LSD estimates followed the expected hemodynamics well. LU and LSD estimation performed worse. Maybe more importantly, we can see from this experiment that MI-LSD does not simply systematically estimate higher sO$_2$ values but indeed estimates plausible values in a wide range.
In a follow-up study of MI-LSD OA oximetry such a hemodynamic challenge experiment could be extended by an oxygen challenge, which would allow a pulse oximetry reference measurement. Other experiments may include a local variation of perfusion e.g.~with adrenaline injections \cite{bunkePhotoacousticImagingMonitoring2021}.

In conclusion, MI-LSD proved highly accurate \emph{in silico} and showed consistently plausible estimates of sO$_2$ \emph{in vivo}. Our preliminary study shows that MI-LSD has the potential to be a robust tool enabling quantitative OA oximetry imaging in humans.

\begin{backmatter}
\bmsection{Funding}
This work has been funded in part by the Swiss National Science Foundation under project no.~ 205320-179038, and the Deutsche Forschungsgemeinschaft (DFG, German Research Foundation) -- 471755457.

\bmsection{Acknowledgments}
Calculations for the Monte Carlo simulations were performed on UBELIX, the HPC cluster at the University of Bern. We thank Adrian Jenk at the Institute of Applied Physics mechanical workshop. For their support of the open-source SIMPA framework we thank Janek Groehl and the Photoacoustics team at the Computer Assisted Medical Interventions division, German Cancer Research Center, Heidelberg.

\bmsection{Disclosures}
The authors declare no conflicts of interest.

\bmsection{Author contributions}
Conceptualization, T.K. and M.F.; Software, T.K. and M.J.; Methodology, Investigation, Formal Analysis, Data curation, Validation, Visualization, Writing – original draft, T.K.; Writing – review \& editing, T.K., M.J. and M.F.; Supervision, Funding acquisition, M.F.

\bmsection{Data and code availability} 
Data underlying the results presented in this paper are fully open-source; available at \href{https://doi.org/10.5281/zenodo.5929161}{doi:10.5281/zenodo.5929161}. The code for the methods described in this paper is fully open-source; available at \href{https://github.com/thkirchner/MI-LSD-in-vivo.git}{github:thkirchner/MI-LSD-in-vivo}. 

\bmsection{Supplemental document}
See Supplement 1 for supporting content.

\end{backmatter}

\bibliography{article}
\end{document}